\documentclass[a4paper,fleqn,usenatbib]{mnras}

\usepackage{mathptmx}
\usepackage[T1]{fontenc}
\usepackage{ae,aecompl}

\usepackage{graphicx}	% Including figure files
\usepackage{amsmath}	% Advanced maths commands
\usepackage{amssymb}	% Extra maths symbols
\usepackage{bm}
\usepackage{threeparttable}
\usepackage{multicol} 
\usepackage{ulem}
\usepackage[title]{appendix}

\title[Radio Suppression by Relativistic Plasma Flows]{Relativistic Fireball Reprise: Radio Suppression at the Onset of Short Magnetar Bursts}

\author[Yamasaki et al.]
{Shotaro Yamasaki\thanks{E-mail: yamasaki@astron.s.u-tokyo.ac.jp}$^{1}$, Shota Kisaka$^{2}$, Toshio Terasawa$^{3,4}$, and Teruaki Enoto$^{5}$
\\
$^{1}$Department of Astronomy, School of Science, The University of Tokyo, Hongo, Bunkyo-ku, Tokyo 113-0033, Japan\\
$^{2}$Department of Physics and Mathematics, Aoyama Gakuin University, Sagamihara, Kanagawa 252-5258, Japan\\
$^{3}$Institute for Cosmic Ray Research, The University of Tokyo,  Kashiwanoha, Kashiwa, Chiba 277-8582, Japan\\
$^{4}$National Astronomical Observatory of Japan,
Osawa, Mitaka, Tokyo 181-8588, Japan\\
$^{5}$The Hakubi Center for Advanced Research and Department of Astronomy, Kyoto University, Kyoto
606-8302, Japan\\
}

\date{Accepted XXX. Received YYY; in original form ZZZ}

\pubyear{2018}

\begin{document}
\label{firstpage}
\pagerange{\pageref{firstpage}--\pageref{lastpage}}
\maketitle

\begin{abstract}
There is growing evidence that a clear distinction between magnetars and radio pulsars may not exist, implying the population of neutron stars that exhibit both radio pulsations and bursting activities could be potentially large. In this situation, new insights into the burst mechanism could be gained by combining the temporal behavior of radio pulsations. We present a general model for radio suppression by relativistic $e^{\pm}$ plasma  outflows  at  the  onset  of  magnetar  flares.  A  sudden  ejection  of  magnetic energy into the magnetosphere would generate a fireball plasma, which is promptly driven to expand at relativistic speed. This would make the plasma cutoff frequency significantly higher than radio frequencies, resulting in the suppression of radio waves. We analytically show that any GHz radio emission arising from the magnetosphere is suppressed for ${\cal O}(100 \,{\rm s})$, depending on the fireball energy. On the other hand, a thermal radiation is expected from the hot spot(s) on the stellar surface created by inflows of dense plasma, which could be the origin of short bursts. Since our hypothesis predicts radio suppression in coincidence with short bursts, this could be  an  indirect  method  to  constrain  the occurrence  rate  of  short  bursts  at  the  faint end that remain undetected by X-ray detectors. Furthermore, we estimate the expected $\mu$sec-scale photospheric gamma-ray emission of plasma outflows. Finally, our model is applied to the radio pulsar with magnetar-like activities, PSR J1119--6127 in light of recent observations. Implications for fast radio bursts and the possibility of plasma lensing are also discussed.
\end{abstract}


%%%%%%%%%%%%%%%%%%%%%%%%%%%%%%%%%%%%%%%%%%%%%%%%%%

%%%%%%%%%%%%%%%%% BODY OF PAPER %%%%%%%%%%%%%%%%%%

\section{Introduction}
\label{sec:intro}

Magnetars \citep{DT1992}, an enigmatic class of highly magnetized neutron star, are known to exhibit flaring activities, broadly classified into ``giant flares"
($10^{44}$--$10^{47}$ erg $\rm s^{-1}$ emitted in several minutes), ``intermediate flares" ($10^{41}$--$10^{43}$ erg $\rm s^{-1}$) or ``short bursts"
($10^{36}$--$10^{41}$ erg ${\rm s^{-1}}$ with duration ranging from a few millisecond to a few second), as well as large and sudden increases (factor of $10$--$1000$ up to $10^{36}$ erg $\rm s^{-1}$, lasting $\lesssim1$ yr) of the persistent emission (``outbursts") which often accompany a variety of anomalies in radiative behaviors (see \citealt{Rea2011,Kaspi2017} for recent reviews).
Among them, short bursts are the most common events, displaying a variety of underlying duty cycles. While some short bursts have clustered distributions in time (``Soft Gamma-ray Repeaters"; SGRs), others do not (``Anomalous X-ray Pulsars"; AXPs).
There is a variety of progenitor models proposed for magnetar flares; some of them are related to an internal instability that leads to the sudden ejection of magnetic energy from the core into magnetosphere \citep{TD1995,TD2001}, while others to an external release of magnetic energy through magnetic reconnections \citep{Lyutikov2003,Gill2010,Yu2012,Parfrey2013,Yu2013}.

\begin{figure*}
 \includegraphics[scale=0.53]{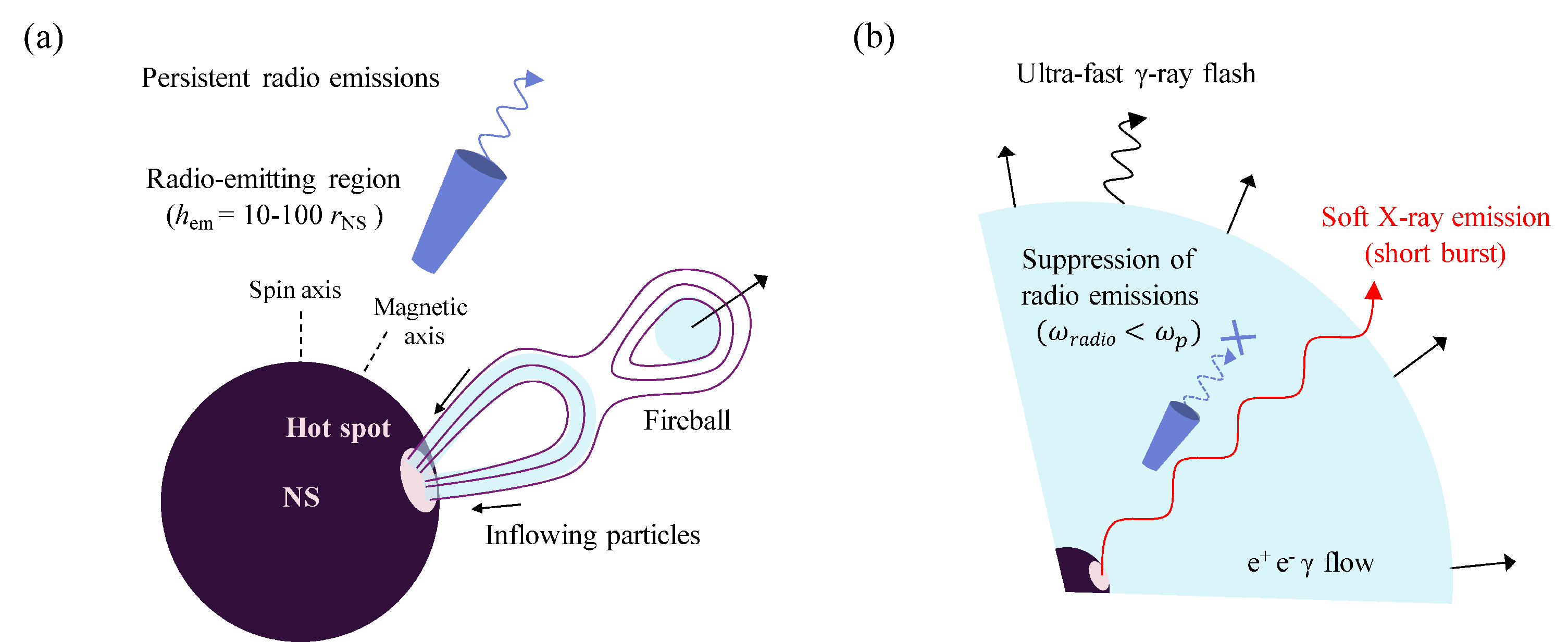}
 \caption{Schematic pictures of the model for short bursts and the radio suppression. (a) At the onset of the fireball release, a fraction of fireball plasma drifts downward along the magnetic loop and heats the surface of the neutron star, creating a hot spot. The pulsed radio emissions are assumed to be generated at altitude $h_{\rm em}=100$--$1000$ km above the neutron star surface. (b) After the fireball expansion, the radio emission site is covered by a dense $e^{\pm}\gamma$ plasma, which makes the plasma frequency significantly higher than radio frequencies, resulting in the suppression of radio emissions. While the hot spot on the stellar surface emanates soft X-ray emissions that are not affected by a plasma flow (the light-bending effect in the vicinity of the stellar surface is neglected for the purpose of presentation).
}
 \label{fig: snapshots}
\end{figure*}

Some magnetars are known to exhibit transient coherent radio pulsations during their burst active phases (e.g., \citealt{Camilo2006,Camilo2007a}), suggesting possible connections to radio pulsars. Meanwhile, magnetar-like short bursts have been discovered from a radio pulsar \citep{Gogus2016,Kennea2016GCN,Younes2016GCN}.
Under such circumstances, any isolated neutron star with magnetar-like activity may potentially have radio pulsations, whereas any ordinary radio pulsar that seems in burst-inactive states can occasionally show bursting activities.
Therefore, the number of neutron stars that exhibit both radio pulsations and bursting activities could be potentially large and considerably increased by future wide-field transient surveys.

Furthermore, \citealt{Archibald2017} have recently reported an observation of magnetar-like short bursts from a  radio pulsar PSR J1119--6127, which coincide in time with the suppression of periodic radio emissions. For each short burst, the persistent $1.4$ GHz radio flux initially gets suppressed, followed by the recovery to its quiescent level on time scale of $\sim10$--$100$ s, which is much longer than the spin period of the pulsar ($P\sim0.4$ s), requiring  new explanations. Given such relationship between radio pulsations and bursting activities,
considering both might enable us to gain new insights into the burst mechanism. In this paper we propose a model for radio suppression mechanism at the onset of short bursts, getting inspirations from the intriguing findings by \citealt{Archibald2017}.
We give a brief overview of our model in what follows (see also figure \ref{fig: snapshots}).

\begin{figure*}
 \includegraphics[scale=0.4]{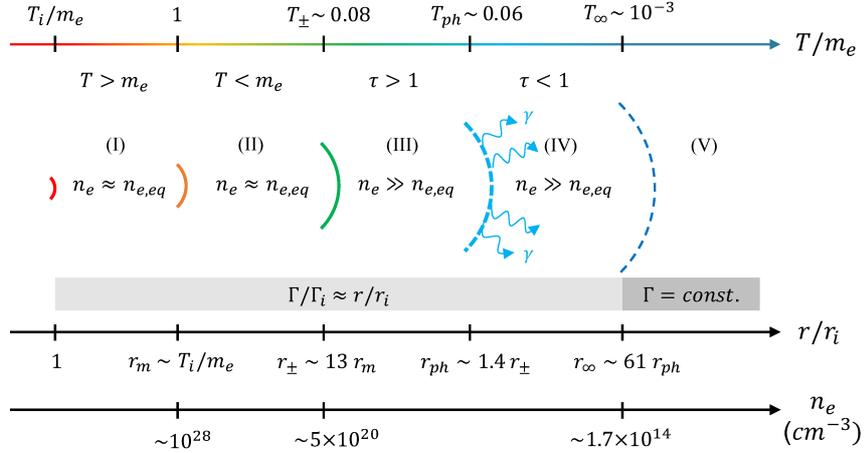}
 \caption{Schematic picture of the fireball evolution with an initial fireball size of $r_i=10^5$ cm.
 The photospheric emissions expected during coasting phase (phase IV) are shown as wavy arrows, which would be observed as ultra-fast gamma-ray flashes.
Variations of $r_i$ do not significantly influence the scaling between adjacent critical radii (only $r_{\pm}/r_m$ weakly depends on $r_i$ as $r_{\pm}/r_m\approx13.4+\log{r_{i,5}}$). See appendix \ref{sec: appendix} for the detailed derivation of each value.}
 \label{fig: FB evolution}
\end{figure*}

We consider a situation that short bursts occur in an isolated neutron star with pulsed radio emissions. Note that we do not necessarily  suppose radio pulsars or magnetars, and thus our model is of wide application. A sudden deposition of the magnetic energy into the magnetosphere may generate an extremely optically thick, compact photoleptonic plasma (so-called ``fireball"). The fireball eruption is expected to occur at the top of the magnetic loop in analogy with solar flares (e.g., \citealt{Lyutikov2006,Masada2010}).
Depending on the pressure balance between the fireball and the magnetic field at the fireball formation site, the fireball is instantaneously driven to expand relativistically by its own internal pressure, acquiring a bulk Lorentz factor of  $\sim10^3$ [see equation (\ref{eq: coasting Lorentz factor}) in appendix \ref{sec: appendix}].

As a consequence of the fireball expansion, the magnetosphere is  covered by a dense $e^{\pm}$ plasma of the fireball, which would make the local plasma cutoff frequency many orders of magnitude larger than radio frequencies ($\sim$ GHz). Pulsar radio emissions are generally considered to be related to the particle acceleration above the polar cap region, which is defined by the last open magnetic field line.  We assume that a similar radio emission mechanism is operated in  bursting neutron stars.
The generation of radio pulses is expected to continue during the fireball expansion, while the surrounding environment becomes dense enough to prohibit the radio emission to propagate. Therefore, whatever the radio emission mechanism is, pulsed radio emissions (if exists) would be inevitably suppressed until the local plasma density sufficiently decreases as the fireball expands. The recovery timescale of the radio emission is determined by the initial fireball properties.

In addition to the outflowing plasma component discussed above, we also consider the inflowing plasma component, which includes a trapped fireball that remains confined to the stellar surface by the closed magnetic fields \citep{TD1995}.
At the onset of the fireball formation, some fraction of the fireball plasma may drift downward along the magnetic loop, bombarding the footpoints (loop base) effectively, which may in turn lead to the formation of a hot spot at the surface of the neutron star. Depending on the energy deposited by the particle inflow, the hot spot emanates soft X-ray emissions lasting $\sim 0.1 $ s, observed as short bursts often seen in the magnetar population. Meanwhile, the outflowing plasma component might also emanate electromagnetic (EM) radiations, which could be observed as a smoking gun.  We consider this possibility and thereby show that the fireball itself produces photospheric emission in hard X-ray to MeV gamma-ray range after entering the optically thin regime, although the detection is challenging due to its extremely short duration ($\sim \mu$s).

This paper is outlined as follows.  In section \ref{sec: Event trigger mechanism}, we describe the triggering mechanism of short bursts, putting an emphasis on the fireball evolution.
We then examine the consequence of the fireball expansion, proposing a model for radio suppression by pair plasma in section \ref{sec: Radio suppression by pair outflows}. The temporal behavior, spectrum and observability of the EM counterpart arising from the fireball photosphere is discussed in section \ref{sec: High-energy counterparts}. Our radio suppression model is applied to the high field radio pulsar with  bursting activities (PSR J1119--6127) in section \ref{subsec: Application to PSR J1119--6127}, and implications for magnetar model of Fast Radio Bursts are presented in section \ref{subsec: FRBs}.
Some discussions on the possibility of plasma lensing and conclusions will be given in section \ref{sec: Conclusions}. The detailed analytic derivation of the fireball evolution is summarized in appendix \ref{sec: appendix}.
Hereafter we often adopt a notation
$Q_x = Q/10^x$ in cgs units and an unit $k_{B}=1=c$ ($k_B$ and $c$ are Boltzmann's constant and speed of light, respectively) regarding the temperature of the fluid (i.e., the temperature $T$ and the electron rest mass energy $m_e$ share the same dimension).

\section{Event trigger mechanism}
\label{sec: Event trigger mechanism}
\subsection{Fireball expansion}
A sudden release of pure energy into a relatively compact volume in the magnetosphere leads to the formation of a radiation-dominated $e^{\pm}$ pair plasma (so-called ``fireball"). We consider a situation that the fireball is not trapped by the magnetic pressure in the magnetosphere.
No sooner is the fireball formed than it frees itself from the confinement of the magnetic pressure
and starts to expand. This is possible depending on the formation height and/or the magnetic field geometry near the surface of the neutron star (e.g., a highly non-dipolar configuration discussed by \citealt{Huang2014a, Huang2014b,Yao2018}).

The fireball is treated as a spherically evolving relativistic fluid composed of $e^{+}e^{-}$ pairs plus $\gamma$ photons (possibly with some baryons as discussed in appendix \ref{sec: appendix}). Photons can be regarded as a relativistic fluid, since they are strongly coupled with pairs due to the extremely optically thick environment.
The conservation of energy and momentum for a steady hydrodynamical flow in spherical symmetry leads to a set of simple scaling laws that govern the radial evolution of the bulk Lorentz factor and temperature
(\citealt{Paczynski1986,Goodman1986}). The bulk Lorentz factor increases linearly with $r$ as
$\Gamma\approx\Gamma_i(r/r_i)$ for $r<r_{\infty}$, where $r_i$ is the initial fireball size and $r_{\infty}$ the saturation radius above which the acceleration of plasma stops and fireball enters a coasting phase with an asymptotic bulk Lorentz factor $\Gamma_{\infty}$. Meanwhile the fireball temperature cools as $T\approx T_i(r/r_i)^{-1}$. The dynamical evolution of fireball is uniquely determined by initial fireball parameters $r_i$, $T_i$ and $\Gamma_i$. For the sake of simplicity, we implicitly assume $\Gamma_i=1$ and different values of $r_i$ and $T_i$ are tried.

In addition to the dynamical evolution, we consider the evolution of the pair number density, taking into account the interactions among pairs and photons (i.e., creation and annihilation).
We denote the number density of fireball electrons (equal to that of positrons) by $n_e$, and hence the net lepton number density $2n_e$.
We assume that the fireball plasma starts to evolve from the equilibrium number density of electrons (and positrons) given as \citep{TD1995}
\begin{eqnarray}
\label{eq: n_e,eq}
n_{e,\rm eq}(T)&\approx&\frac{2}{(2\pi)^{3/2}}\,\lambda_C^{-3}\,\left(T/m_e\right)^{3/2}\,e^{-m_e/T}\nonumber \\%
&\sim& 10^{28} \ \left(T/m_e\right)^{3/2}\,e^{-m_e/T} \ \ {\rm cm^{-3}},
\end{eqnarray}
where $\lambda_C$ is the Compton length. The quantum effects under the magnetic field higher than the critical field strength $B_Q\equiv m_e^2c^3/(e\hbar )\sim4.4\times10^{13}$ G may change the equilibrium  number density by a factor of $\sim B/B_Q$ \citep{Harding2006}, but barely affect the result.
To summarize, the evolution of electron (positron) number density $n_e$ is characterized by several critical radii that determine the physical properties of the fireball \citep{Grimsrud1998,Iwamoto2002,Li2008}:

\begin{table}
\centering
\caption{Description of critical radii that control the fireball evolution.}
\label{tab: critical radii}
\begin{tabular}{ll}
\hline
\hline
$r_i \cdots$    & Initial fireball size   \\
$r_m\cdots$     & Electron temperature radius (at which $T= m_e$) \\
$r_{\pm}\cdots$ & Pair equilibrium breakup radius   \\
$r_{ph}\cdots$  & Photospheric radius (at which $\tau\sim1$)  \\
\multicolumn{1}{l|}{$r_{\infty}\cdots$} & \multicolumn{1}{l|}{Coasting radius} \\ \hline
\end{tabular}
\end{table}

\begin{description}
\item[{\bf (I)--(II)}] The initial fireball is at rest in pair equilibrium due to its high temperature with its size $r=r_i$. It immediately expands and cools down to the electron rest mass energy at $r=r_m$, and then the number density of pairs begins to deviate from the equilibrium number density at $r=r_{\pm}$.
\item[{\bf (III)}] The pair annihilation dominates the pair process since the number of pair-creating high-energy photons decreases as the fireball cools. Eventually, the fireball reaches the photospheric radius $r=r_{ph}$ at which the optical depth to electron scattering becomes an order of unity.
\item[{\bf (IV)}] When the fireball becomes optically thin, photons begin to leak freely out of the photosphere. However, they still continues to supply the radiation energy to pairs, which accelerates pairs up to the coasting radius $r=r_{\infty}$.
\item[{\bf (V)}] The photons cease to inject the radiation energy to pairs, and the fireball begins to freely coast at constant speed $\Gamma=\Gamma_{\infty}$. The pair annihilation no longer occurs due to the small number density. Therefore, the total number of pairs conserves and the pair density evolves as $\propto r^{-2}$.
\end{description}

\begin{figure*}
 \includegraphics[width=\linewidth]{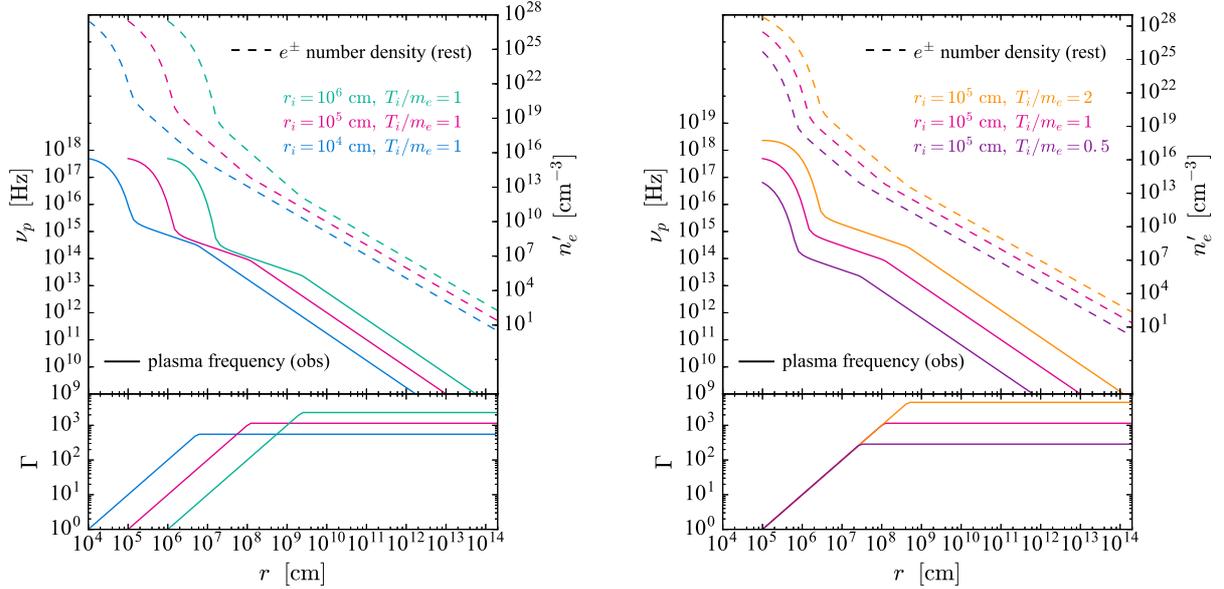}
 \caption{The evolution of electron number density in the plasma rest frame (upper panels, dashed lines), plasma frequency in the observer frame (upper panels, solid lines) and bulk Lorentz factor of fireball plasma (lower panels) for different initial sizes (left) and temperatures (right).  Equations (\ref{eq: phase I}), (\ref{eq: phase II}), (\ref{eq: phase III to IV})  and (\ref{eq: phase V}) are combined to describe the overall evolution of $n_e^{\prime}$, which is then translated into $\nu_p$ by using equation (\ref{eq: plasma freq.}) .
 The initial radius and temperature can be converted into the total fireball energy by using equation (\ref{eq: E_fb}).}
 \label{fig: nu_p & n_e evolution}
\end{figure*}

Finally, we obtain the radial evolution of the fireball from the analytic estimates as shown in appendix \ref{sec: appendix}, which is also summarized in figure \ref{fig: FB evolution}.
Hereafter, we often relate initial fireball parameters with the total fireball energy by
\begin{eqnarray}
\label{eq: E_fb}
E_{\rm fb}=aT_i^4r_i^3\sim10^{40}\,r_{i,5}^3\, \left(\frac{T_i}{m_e}\right)^4 \ {\rm erg}.
\end{eqnarray}Although we adopt an initial fireball size of $r_i=10^5$ cm as a reference assuming a typical total energy for short bursts $\sim 10^{40} {\ \rm erg}$, the results on radii presented in figure \ref{fig: FB evolution} can be easily scaled to other values of $r_i$, since only $r_{\pm}/r_m$ is weakly dependent on $r_i$ [see equation (\ref{eq: analytic r_pm/r_m})]. These results will be used to estimate the time scale for radio suppression and recovery in section \ref{sec: Radio suppression by pair outflows}.

\subsection{Hot spot formation}
\label{sec: Hot spot formation}
At the onset of a fireball eruption, some fraction of the fireball plasma may be left at the looptop and stream downward along the magnetic loop. These energetic particles immediately bombard the loop base, which leads to create an inhomogeneity in temperature (``hot spot") at the stellar surface  which might be comparable in size to the initial fireball. Although the formation of multiple hot spots is possible depending on configuration and size of magnetic loops, here we consider a single hot spot as a whole for simplicity.
We assume that a fraction of the initial fireball energy $E_{\rm fb}$ is
converted via the bombardment into the energy of the hot spot $E_{\rm rad}$, which is immediately radiated away by the blackbody emissions.
If the hot spot with radius $r_{\rm spot}$ cools by radiating thermal emissions with temperature $T_{\rm spot}$, the phenomenological duration of the thermal radiation from the hot spot may be estimated as
\begin{eqnarray}
\label{eq: duration of the hot spot emission}
    \Delta t_{\rm rad} &\approx& \frac{E_{\rm rad}}{\sigma_{SB} T_{\rm spot}^4 4\pi r_{\rm spot}^2} \nonumber \\%
    &\sim& 80 \ \ \,E_{\rm rad,38}\,r_{{\rm spot},5}^{-2}\left(\frac{T_{\rm spot}}{10\ {\rm keV}}\right)^{-4} \ {\rm ms},
\end{eqnarray}
where $\sigma_{SB}$ is the Stephan--Boltzmann constant. Here we adopt a typical blackbody temperature $\sim10$ keV for $E_{\rm rad}\sim10^{38}$ erg burst (e.g., 2016 July bursts of PSR J1119--6127, \citealt{Gogus2016}).  While the hot spot size $r_{\rm spot}$ could presumably be related to the initial fireball size $r_i$ as $r_{\rm spot}\approx r_i$ as far as $r_{i}\lesssim10^6$ cm, the hot spot temperature is expected to linearly scale with the total energy of radiation as $E_{\rm rad}\propto T_{\rm spot}^4$. This implies that the duration of the flare estimated above would be almost constant for a wide range of observed radiation energy $E_{\rm rad}=10^{36}$--$10^{41}$ erg, which is broadly consistent with the peak $\sim100$ ms in the duration distribution of short bursts \citep{Kaspi2017}.
The efficiency of the surface radiation $E_{\rm rad}/E_{\rm fb}(\lesssim 1)$ is highly uncertain due to the lack of knowledge on the energy dissipation process at the neutron star surface.
Given the smaller size and the higher temperature of the hot spot compared to the whole stellar surface, the resulting emission should be observed as pulsed emissions, whose pulsed fraction could be either small or large, depending on the geometry relative to the observer.

\section{Radio suppression by pair outflows}
\label{sec: Radio suppression by pair outflows}
\subsection{Persistent radio emissions from magnetars}
Up to the present, coherent radio pulsations have been detected from only four magnetars, all of which are transient, emerging in coincidence with X-ray outbursts \citep{Camilo2007a,Camilo2007b,Levin2010,Kaspi2014}.
The high-energy radiation of magnetars is generally considered to be powered by the magnetic energy, since its characteristic quiescent X-ray luminosity $10^{34}$--$10^{36}\ \rm{erg \  s^{-1}}$ \citep{Rea2011} is in  excess of the rotational energy loss rate due to the magnetic braking (so-called ``spin-down luminosity'')
$L_{\rm sd}=3.9\times10^{35} \,B_{14}^2 R_6^6 \,P^{-4} \ {\rm erg \ s^{-1}}$ $=10^{32}$--$10^{34} \ \rm{ erg \ s^{-1}}$ where $R$ is the neutron star radius and $P=P/({\rm 1 \ s})$.
On the other hand, radio pulsations from magnetars are normally faint (well below $L_{\rm sd}$) and might be powered by the rotational energy (e.g., \citealt{Szary2015}).
 Namely, coherent radio emissions from magnetars may be generated by the relativistic plasma flow accelerated outwards along the open magnetic field lines with an emission altitude of  $h_{\rm em}=10$--$100\, R$ above the stellar surface, as is likely the case for conventional radio pulsars.
In that case, the persistent radio emission from magnetars should be beamed as radio pulsars although the beam size could be temporarily changed by the bursting activity (e.g., \citealt{Beloborodov2008,Szary2015}). The transient, low-efficiency, and anisotropic nature of radio emission all indicate that the detection rate of pulsed radio emissions of magnetars could be low, which is consistent with observations.
Given the circumstantial evidence above, we assume that a similar radio emission mechanism operates both in magnetars and radio pulsars.

\subsection{Radio suppression and recovery}
\label{subsec: Radio suppression and recovery}
When the fireball begins to  expand, the surrounding environment of the radio-emitting region is covered by a fireball plasma.  In general, the motion of a charged particle is strongly confined along the magnetic field line inside the magnetosphere, and the fireball plasma cannot interact with particles that are responsible for radio emissions. For this reason, we can reasonably assume that the generation of radio emissions continues during the fireball expansion.

As a consequence of an expanding plasma flow, however, it is expected that any radio emission at $\sim$ GHz frequencies arising in the magnetosphere would suffer from the absorption. Here we consider a radio suppression due to the damping of waves by the pair plasma as a relevant absorption process.
The plasma frequency of the fireball outflow measured in the observer frame is defined as
\begin{eqnarray}
\label{eq: plasma freq.}
\nu_p=\frac{\Gamma}{2\pi}\sqrt{\frac{4\pi n_e^{\prime} e^2}{m_e}}\sim9.0\times10^3 \ \Gamma\, {n_e^{\prime}}^{1/2} \ {\rm Hz},
\end{eqnarray}
below which the radiation in general cannot propagate through the medium. Here $\Gamma$ is the bulk Lorentz factor of the fireball and $n_e^{\prime}$ the electron number density measured in the comoving frame of the fireball plasma.
The evolution of $n_e^{\prime}$ and $\nu_p$ under different initial conditions is shown in figure \ref{fig: nu_p & n_e evolution}. One can see that the fireball is in the coasting phase [phase (V)] when the plasma frequency reaches observing radio frequencies $\sim$GHz. At this stage ($r\gg r_{\infty}$), the bulk Lorentz factor of the plasma stays constant $\Gamma=\Gamma_{\infty}$, and the electron number density decreases as $n_e^{\prime}=n_e^{\prime}(r_{\infty})(r/r_{\infty})^{-2}$.
Substituting these to equation (\ref{eq: plasma freq.}), we can express the the plasma frequency at phase (V) as a function of time ($t\sim r/c$):
\begin{eqnarray}
\label{eq: n_e at t = t_rec}
\nu_p(t)\approx 4.7\times10^{11}  \ r_{i,5}^{5/8} \left(\frac{T_i}{m_e}\right)^2\,t^{-1} \ \ {\rm Hz}
\end{eqnarray}
Then, the characteristic time scale for the recovery of the pulsed radio emissions from suppression by the pair plasma defined by $\nu_p(\tau_{\rm rec})=\nu\sim$ GHz is estimated as
\begin{eqnarray}
\label{eq: t_rec}
\tau_{\rm rec}\approx 4.7\times10^{2}  \ \ r_{i,5}^{-7/8} \left(\frac{E_{\rm fb}}{10^{40} {\rm \ erg}} \right)^{1/2} \,\nu_{9}^{-1}\ \ \,{\rm s},
\end{eqnarray}
where we have introduced the total fireball energy $E_{\rm fb}$ by using equation (\ref{eq: E_fb}) .
From the observational perspective, $\tau_{\rm rec}$ is directly obtained by examining a radio light curve and $r_i$ might be related to the observed hot spot size, both of which allow us to estimate the total fireball energy $E_{\rm fb}$.
Although a baryon-free fireball is implicitly assumed here,
the possible baryon contamination at the time of fireball formation does not significantly change our result. We estimate that the heavy baryon loading would increase $\nu_p$ only by a factor of $\lesssim$ a few (see Appendix \ref{subsec: Baryon loaded fireball}).
One possible uncertainty in our model is the assumption on the isotropic expansion of fireball. Given the relatively large initial Lorentz factor of the fireball (e.g., $\Gamma_i\gtrsim3$), it is expected that the fireball may expand in a highly anisotropic manner; this could result either in  delayed radio suppression or none whatsoever. We leave the exploration of this possibility for a future work.

Intriguingly, the spectrum of pulsed radio emissions from four magnetars is known to be flat across wide frequency ranges (typically $1$--$100$ GHz, \citealt{Kaspi2017}).
A natural consequence of this is that the recovery of radio emissions from the complete suppression would take place gradually (not abruptly) while the plasma frequency passes through the spectrum energy ranges of magnetars. Given the observing frequency band $\nu\in[\nu_1,\nu_2]$ that is contained by the flat spectrum frequency ranges of magnetars, the radio emissions are completely suppressed  ($\nu_p\gg\nu_2$) at early times, and then followed by the subsequent partial recovery phase ($\nu_p\in[\nu_1,\nu_2]$), whose time scale could be the same order as $\tau_{\rm rec}$ for a band width $\Delta \nu\sim\nu$.

\section{High-energy counterparts}
\label{sec: High-energy counterparts}
\begin{figure}
 \includegraphics[width=\linewidth]{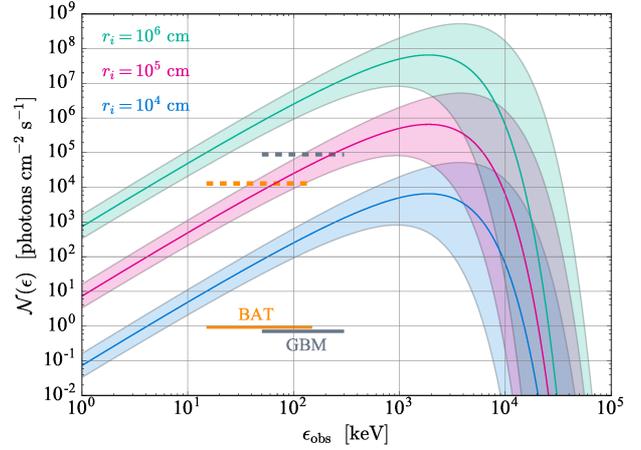}
 \caption{The observed spectrum of the photospheric emissions with different initial fireball size (same as figure \ref{fig: nu_p & n_e evolution}).
 Results for the initial temperature of $T_i=m_e$ are shown in solid lines with shaded regions corresponding to a factor of two higher and lower temperature. The initial radius and temperature can be converted into the total fireball energy by using equation (\ref{eq: E_fb}).
 The trigger sensitivities of BAT and GBM are shown in solid horizontal lines. The exposure-corrected sensitivities estimated using equation (\ref{eq: apparent sensitivity}) are also shown in dashed horizontal lines.
}
 \label{fig: photospheric emission}
\end{figure}

In section \ref{sec: Hot spot formation}, we interpret observed short bursts as thermal emissions from the hot spot that is generated by particles inflowing toward the stellar surface. On the other hand, the expanding fireball is also expected to emanate radiations at times when it becomes optically thin.
Detecting such a signature of the expanding fireball would be useful to examine the validity of our scenario presented in section \ref{sec: Radio suppression by pair outflows}.
Here we consider EM wave signatures from expanding fireball component, which should be distinguished from short bursts originated from inflowing component of the fireball plasma.
While the rest frame temperature of the fireball decreases monotonically with increasing radius ($T\propto r^{-1}$), the emissions from the photosphere would be a blackbody with the Doppler-boosted temperature
\begin{eqnarray}
T_{\rm obs}=\frac{T}{\Gamma(1-\beta\mu)}\equiv {\cal D}T,
\end{eqnarray}
where $\beta\mu$ is the projection of the three velocity onto the line-of-site, and $\cal{D}$ being the Doppler factor. Because of the blue-shifted temperature (by a factor of ${\cal D}\sim2\Gamma$--$\Gamma$ for $\theta=0$--$1/\Gamma$), the peak energy ranges relatively wider. The observed photospheric emissions would peak at
$\epsilon_{\rm obs, peak}\sim2\Gamma_{\rm ph}T_{\rm ph}\approx 2\Gamma_iT_i$,
where we have used $\Gamma T=const$.
The flux per energy interval received by a distant observer is given as
\begin{eqnarray}
{\cal N}\,(\epsilon_{\rm obs})&=&\frac{4\pi\epsilon_{\rm obs}^3}{h^3c^2}\int_{-1}^{1}\frac{\mu d\mu}{\exp{\left(\epsilon_{\rm obs}/T_{\rm obs}\right)}-1}\nonumber \\%
&\approx&\frac{4\pi\epsilon_{\rm obs}^2}{h^3c^2}\frac{T}{\Gamma}\left\{-\ln{\left[1-\exp{\left(-\frac{\epsilon_{\rm obs}}{2\Gamma T}\right)}\right]}\right\}.
\end{eqnarray}
Using $T/\Gamma\propto r^{-2}$ and $\Gamma T =const$, the observed spectrum at $d=10$ kpc is presented in figure \ref{fig: photospheric emission}.
The observed duration of the photospheric emission is expressed as
\begin{eqnarray}
\delta t_{\rm obs}=\delta t(1-\beta \mu)=\delta t^{\prime} \Gamma (1-\beta \mu)={\cal D}^{-1}\delta t^{\prime},
\end{eqnarray}
where $\delta t^{\prime}$ and $\delta t$ denote the time interval of emitted two photons in the source comoving frame and in the observer frame, respectively (hence $\delta t=\Gamma\delta t^{\prime}$). We can see that the arrival time difference of two photons is affected by both the Lorentz-boost and the purely geometrical effect. The burst lasts during phase (IV) in the source rest frame $\delta t^{\prime}\sim r_{\infty}/c\lesssim10^{-2}$ s. Therefore, the observed duration is extremely short: $\delta t_{\rm obs}\sim\delta t^{\prime}/\Gamma_{\infty}\lesssim10^{-5}$ s.

Let ${\cal N}_{\rm lim}$ the onboard trigger sensitivity of a gamma-ray detector at observing energy $\epsilon_{\rm obs}$ with sampling time window $t_{\rm exp}$. In general, detectors are optimized for transients with duration longer than the minimum sampling time window $\sim{\cal O}$(ms). In the case of ultra-fast gamma-ray flashes with typical duration $\delta t_{\rm obs}\sim{\cal O}(\mu{\rm s})$ ($\ll t_{\rm exp}$), however, the trigger threshold should be corrected for the sampling time window as
\begin{eqnarray}
\label{eq: apparent sensitivity}
    {\cal N}_{\rm lim}^{\rm app}={\cal N}_{\rm lim}\times \frac{t_{\rm exp}}{\delta t_{\rm obs}},
\end{eqnarray}
where ${\cal N}_{\rm lim}^{\rm app}$ is the apparent trigger sensitivity. Adopting ${\cal N}_{\rm lim}\sim0.21\,(t_{\rm exp}/1 \, {\rm s})^{-1/2} \ {\rm cm^{-2} \ s^{-1}}$(at  $\epsilon_{\rm obs}=15$--$150$ keV, assuming mean photon energy of $60$ keV) with the minimum sampling time window $t_{\rm exp}=4$ ms for the {\it Swift}
Burst Alert Telescope (BAT, \citealt{Barthelmy2005}), and ${\cal N}_{\rm lim}\sim0.71\,(t_{\rm exp}/1 \, {\rm s})^{-1/2} \ {\rm cm^{-2} \ s^{-1}}$ (at $\epsilon_{\rm obs}=50$--$300$ keV) with $t_{\rm exp}=16$ ms for the {\it Fermi} Gamma-ray Burst Monitor (GBM, \citealt{Meegan2009}), we estimate the apparent trigger sensitivity as shown in figure \ref{fig: photospheric emission}. It can be seen that the event triggering onboard may be challenging unless the initial fireball size is relatively large ($\sim10^6$ cm).
Still, it is particularly interesting to search the archival X-ray or gamma-ray data for photon events clustering within a sub-millisecond time window.  Conversely, the ultra-fast photospheric emission discussed above is observable when the fireball energy is sufficiently high, and this could be the origin of initial gamma-ray spikes of giant flares (e.g., \citealt{TD2001,Lyutikov2006}).

Another possible EM counterpart is afterglow emissions. In the case of giant flares, radio afterglow emissions have been detected from two sources \citep{Frail1998,Gaensler2005}. For instance, late-time observations of the radio afterglow from SGR 1806-20 set a constraint on the total kinetic energy of ejecta  $\sim10^{44}$ erg \citep{Granot2006}, which is in agreement with the fireball model if we take into account a heavy baryon loading \citep{Li2008} or possible magnetic loading \citep{Lyutikov2006}.
However, at relatively lower energies considered here ($E_{\rm fb}\lesssim10^{40}$ erg), the expected kinetic energy of fireball plasma in the coasting phase falls far short of the energy required to power observable radio afterglows.

\section{Applications}
\label{sec: Applications}
\subsection{PSR J1119--6127}
\label{subsec: Application to PSR J1119--6127}
The radio pulsar PSR J1119--6127 was first discovered by Parkes 1.4 GHz pulsar survey with a spin period $P\sim0.4$ s, spin-down rate $\dot{P}\sim4\times10^{12} \ {\rm s \ s^{-1}}$ and spin-down luminosity of $L_{\rm sd}\sim2.3\times10^{36} \ {\rm erg \ s^{-1}}$ at $8.4$ kpc \citep{Camilo2000}. These spin-down parameters indicate that PSR J1119--6127 is relatively young (with characteristic age $1.9$ kyr) and its surface dipole magnetic field strength ($B\sim4.1\times10^{13}$ G) is close to the critical field strength. Several short bursts were detected on 2016 July 27--28 by {\it Swift}--BAT and {\it Fermi}--GBM \citep{Gogus2016,Kennea2016GCN,Younes2016GCN} with a large flux increase in the soft X-ray band (outburst, \citealt{Archibald2016}), after which the radio pulsations became undetectable for two weeks and re-activated again \citep{Burgay2016ATel} with a change seen in the radio pulse profile \citep{Majid2017}.

More recently, \citealt{Archibald2017} has reported the detection of three short bursts on 30 August 2016 (with an average energy of $10^{37}$ erg emitted within a few seconds) that coincide in time with the suppression of persistent radio fluxes.
The burst spectrum is fitted with a blackbody with peak temperature $\sim2$ keV, and the radiative area at $d=8.4$ kpc distance is estimated to be about $1$ km.
As a possible explanation for this, they interpret the short bursts as a thermal emission from the magnetically confined fireball \citep{TD1995}, and consider a leakage of a pair plasma from the trapped fireball into the particle-accelerating region, which would shield the electric field, resulting in the suppression of radio emissions. In that case, however, the cessation of particle acceleration should occur abruptly (with at most light-crossing time of the radio-emitting region $\sim$ ms) rather than continuously once the sufficient number of $e^{\pm}$ pairs is supplied \citep{Lyubarsky2009,Timokhin2010,Timokhin2013,Kisaka2016}, which seems contradictory with their interpretation of the ``gradual" radio recovery (the radio light curve is fitted with an exponential recovery model with a typical time scale $\sim70$ s). Moreover, the formation of a fireball with its surface temperature ($\sim2$ keV), which is about three orders of magnitude lower than the internal temperature $m_e\sim 511$ keV, is unlikely.

Alternatively, we apply our general radio suppression model to PSR J1119--6127 below.
The time scale for the radio recovery $\lesssim100$ s could be accounted for by adopting fireball parameters of $r_i\sim10^5$ cm and $T_i\sim 0.5\,m_e$ from equation (\ref{eq: t_rec}). This translates into an initial fireball energy of $E_{\rm fb}=aT_i^{4}r_i^3\sim10^{38}$ erg, which is sufficiently large to generate short bursts from surface hot spots (i.e., with an efficiency $E_{\rm rad}/E_{\rm fb}\sim 0.1$). Substituting the observed blackbody temperature of the short bursts $\sim2$ keV and the size of the radiating region $r_{\rm spot}=10^{5}$ cm \citep{Archibald2017} into equation (\ref{eq: duration of the hot spot emission}), we estimate the duration of the thermal emissions to be  $\Delta t_{\rm rad}
\sim 5 \ \ \,E_{\rm rad,37}\,r_{{\rm spot},5}^{-2}\left(T_{\rm spot}/2\ {\rm keV}\right)^{-4} \ {\rm sec}$, where we have assumed that about 10\% of the fireball energy ($E_{\rm fb}=10^{38}$ erg) is converted into radiation energy of the surface hot spot.
This roughly agrees with the observed duration $\sim2$--$4$ s and the total radiation energy $\sim10^{37}$ erg of three short bursts on August 30th 2016 \citep{Archibald2017}.

Furthermore, the problem of the magnetic confinement is easily solved if we assume a moderate formation height of the initial fireball.  Given a dipole magnetic field $B\propto r^{-3},$
the magnetic pressure at an altitude $h_{\rm fb}$ above the stellar surface is
$P_B= B^2/(8\pi)\sim6.4\times10^{25} (B_p/4\times10^{13} \ {\rm G})^2\,h_{{\rm fb},6}^{-6}\ {\rm erg \ cm^{-3}}$, whereas the total pressure of the fireball with initial temperature $T_i$ is
$P=P_e+P_r=11/4\,P_r\sim10^{24}\,(T_i/0.5\,m_e)^4\ {\rm erg \ cm^{-3}}$, where $P_e$ and $P_r$ are plasma pressure and radiation pressure, respectively.  The fireball therefore can escape from the magnetic trapping (i.e., $P\gtrsim P_B$) when the initial condition $h_{\rm fb}\gtrsim20$ km is satisfied.  This critical height could be even smaller if the magnetic field is dominated by the higher multipoles close to the neutron star surface.
The non-detection of the high-energy counterparts associated with these short bursts is marginally consistent with the expected burst flux shown in figure \ref{fig: photospheric emission}.

We also apply our model to the two-week disappearance of radio pulsations after the 2016 July 27--28th short bursts \citep{Burgay2016ATel}. Given the two-week radio suppression caused by a single burst, the initial fireball energy must exceed $10^{48}$ erg, which is comparable to that of giant flares and thus seems unlikely. Instead, a viable scenario is that a series of short bursts that are too faint to be detectable occurred repeatedly or in succession during the initial outburst in July 2016, which suppressed the radio pulsation continuously for up to two weeks.

\subsection{Fast radio bursts}
\label{subsec: FRBs}
Fast radio Bursts (FRBs) are short-duration ($\sim$msec), coherent ($\sim$GHz), bright ($\sim$Jy) radio flashes with $\sim10^{38}$--$10^{40}$ erg of energy per burst (\citealt{Lorimer2007,Petroff2016}). Although their origin remains unknown, the large dispersion measure (defined by the number density of free electrons integrated over the line-of-sight) of FRBs greatly exceeds that for Galactic pulsars, suggesting their extragalactic origins.
Only one FRB (FRB 121102) is confirmed to repeat more than 100 times at least. The detection of persistent radio counterpart with luminosity $\sim10^{39}$ erg $\rm s^{-1}$ \citep{Chatterjee2017} leads to the firm localization of the host galaxy at redshift $\sim0.19$. The possible connection between FRB 121102 and a young neutron star born after either supernova (e.g., \citealt{Kashiyama2017,Metzger2017,Beloborodov2017,Margalit2018}) or binary neutron star merger \citep{Yamasaki2017} has been discussed in the literature.

\begin{figure*}
 \includegraphics[scale=0.45]{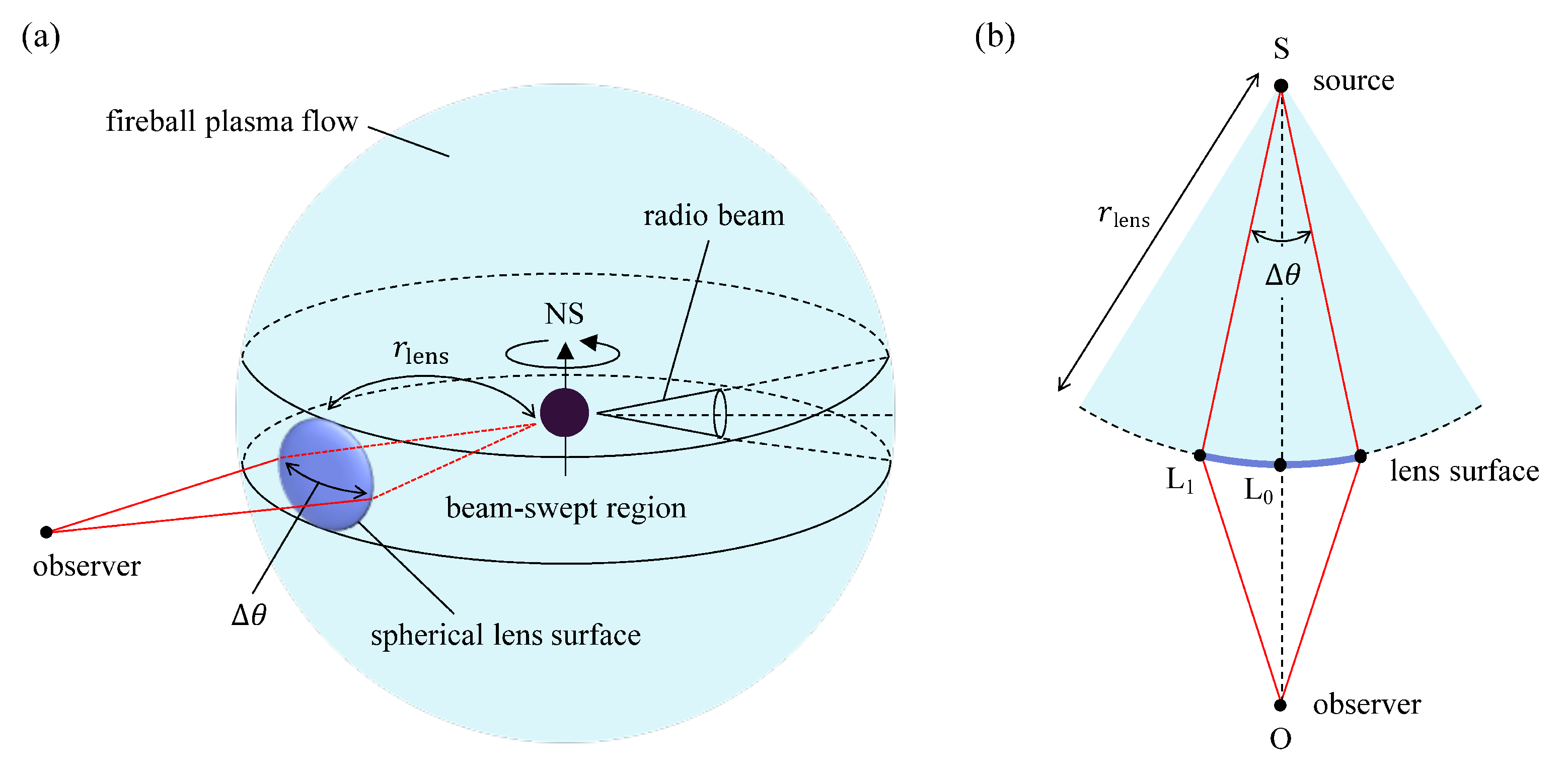}
 \caption{Geometries of scattering for a fireball plasma lens. (a) The observer is assumed to be located in the beam-plane which is perpendicular to the spin axis. The light blue sphere represents the fireball plasma that is responsible for the plasma lensing. We assume a circular radio beam and its projection onto the lens plane is shown as an equatorial belt (``beam-swept region"). Since the timescale for lensing must be much longer than the spin period in order to avoid the radio suppression, this beam-swept region can be regarded as being constantly radiating radio emissions. A light ray which intersects the spherical lens surface by the pitch angle $\Delta\theta$ is assumed to be deflected toward the observer.
 (b) Same as (a) but viewed in the beam-plane.
}
 \label{fig: lens}
\end{figure*}

Provided that the neutron star activity is directly responsible for the FRB generation, one possible trigger mechanism is magnetar flares (e.g., \citealt{Popov2007,Lyubarsky2014,Kulkarni2015,Pen2015,Katz2016,Murase2016,Beloborodov2017}).
The minimum requirement for these models is that the flare energy must exceeds that of typical FRBs ($10^{38}$--$10^{40}$ erg).
As shown in the left panel of figure \ref{fig: nu_p & n_e evolution}, our radio suppression model indicates that the $\sim$GHz radio emission associated with an expanding plasma must originate at the distance $r_{\rm em}\gtrsim 10^{13} \ r_{i,5}^{5/8}(T_i/m_e)^2\nu_{9}^{-1}$ cm from the neutron star under the most optimistic FRB efficiency of order unity.
This in turn rules out any FRB models that predicts the radio wave generation in the vicinity of the neutron stars (i.e., inside the magnetosphere $\lesssim10^{10}$ cm).
In contrast, our model prefers scenarios in which FRBs are generated outside the magnetosphere, e.g., by the maser emission due to the termination shock \citep{Lyubarsky2014} or the internal shocks \citep{Beloborodov2017}.
In such cases, our model predicts that an FRB should occur at least $r_{\rm em}/c\sim100$ s behind the giant flare.

\section{Discussion and Conclusions}
\label{sec: Conclusions}

In this paper, we investigated relativistically expanding fireball plasma as a possible
origin both for short bursts and simultaneous suppression of the persistent radio emission from bursting neutron stars.

A sudden release of the magnetic energy inside the magnetosphere generates a fireball composed of $e^{\pm}$ plasma and radiation. Under the condition that the radiation pressure of the fireball exceeds the magnetic pressure, the fireball begins to expand at relativistic speed, covering the magnetosphere with dense $e^{\pm}$ plasma. This would lead the plasma frequency to greatly exceed the radio frequency in the rest frame of the plasma, resulting in suppression of the persistent (pulsed) radio emission of bursting neutron stars.
We analytically derive the radial evolution of the plasma number density, and estimate the characteristic time scale for the recovery of radio suppression to be $\sim100$ s for a $\sim10^{40}$ erg fireball.
On the other hand, some fraction of the fireball plasma may heat the stellar surface via the particle bombardment, creating hot spots. The particle energy is converted to the hot spot and immediately radiated away as a thermal emission, which can give rise to a short burst with typical duration of $\sim100$ ms. The ultra-fast gamma-ray flashes
from the expanding fireball photosphere are expected as a smoking gun, although the onboard detection by current gamma-ray telescopes might be challenging due the extremely short duration $\sim\mu$s.

Then we applied our hypothesis to the radio pulsar PSR J1119--6127 with magnetar-like short bursts. The observed radio suppression timescale $\lesssim100$ s is well explained by fireball parameters of $r_i\sim10^5$ cm and $T_i\sim 0.5\,m_e$, corresponding to an initial fireball energy of $aT_i^{4}r_i^3\sim10^{38}$ erg. This also permits $\sim10^{37}$ erg short bursts at $\sim2$ keV. The expected gamma-ray counterpart has not been reported yet for J1119--6127, presumably due to the detector's sensitivity limit. However, an archival search for clustered photon events in an extremely short time window contemporaneous with the time of short burst detection would be highly intriguing.
Our model can naturally explain the J1119 observations well, and this might be one of the causes of the nulling and mode-changing commonly seen in radio pulsars.
The implications for FRBs are also discussed. Provided that FRBs are generated by a plasma outflow that is also responsible for magnetar flares, we argue that the radio emission must be produced at $r\gtrsim10^{13}$ cm from the neutron star, which is a minimum requirement to avoid the absorption by the plasma cutoff effect.

Our fireball model naturally explains the radio suppression associated with bursting activities in J1119--6127. However, it should be noted that a similar radio switch-off phenomenon is commonly observed in radio pulsars: nulling and mode changing. The global change in the magnetosphere is proposed as a possible mechanism for them (e.g., \citealt{Wang2007,Timokhin2010b}). Indeed, since the duration of nulling and mode changing varies from one or two rotations to even days, one cannot exclude the possibility that they had played a role for 100 s radio suppression seen in J1119--6127. The change in the mean radio pulse profile of J1119--6127 also resembles those seen in mode changing pulsars \citep{Archibald2017}. Therefore, instead of ruling out this possibility, we only point out that our model could be related to nulling and/or mode changing.  In particular, the radio suppression by a fireball with its energy sufficiently low to remain undetected in X-ray bands would manifest as nulling. This could be the origin of some short-duration nulling events.

In most part of this paper, we have treated the fireball plasma as radio absorber. After the recovery of radio pulses, however, it might also work as a plasma lens that leads to the light amplification,  depending on the geometry and configuration of radio pulses. Let us consider the simplified geometry shown in figure \ref{fig: lens} (a), when the pulsar rotation axis is perpendicular to its magnetic momentum. The observer is assumed to be located in the beam-plane which is perpendicular to the rotation axis, and ray paths in this plane are modelled as shown in figure \ref{fig: lens} (b).
We consider a radio pulse that leaves the source at $t=t_{\rm p}\,(>\tau_{\rm rec})$ and reaches the lensing point at $t=t_{\rm lens}$, corresponding to a travel distance of $r_{\rm lens}= c\,(t_{\rm lens}-t_{\rm p})$. For simplicity, we assume the light refraction takes place at $r=r_{\rm lens}$ with a pitch angle $\Delta \theta$ as in the 1D thin lens model.
In general, the phase of an EM wave is given by the contributions of geometrical and dispersive (group) time delays over a whole ray path:
\begin{eqnarray}
\Phi(u)=\Phi_{g}(u)+\Phi_{\rm DM}(u),
\end{eqnarray}
where $u$ is the transverse coordinate in the lens plane. A minimum requirement for the strong lensing is $\Delta \Phi_{g}\sim\Delta \Phi_{\rm DM}$ so that $\nabla \Phi=0$ at some spacial scales \citep{Main2018}.

\begin{figure}
 \includegraphics[width=\linewidth]{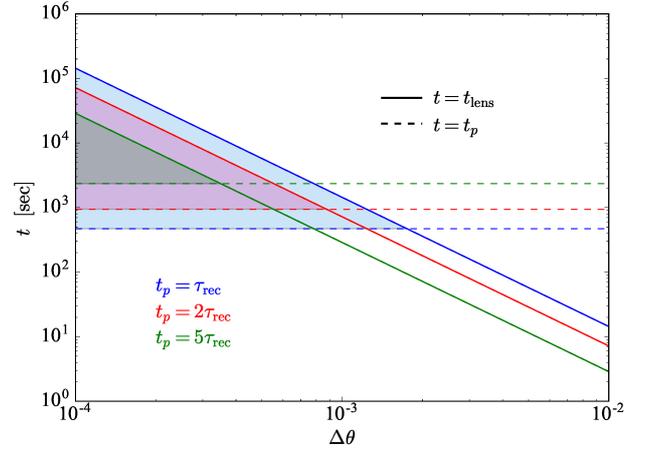}
 \caption{Constraints on the parameter space that might produce strong plasma lensing event: the elapsed time $t$ since the burst onset and lensing pitch angle $\Delta\theta$ with assumptions of $t_p = \tau_{\rm rec},\,2\tau_{\rm rec},\,5\tau_{\rm rec}$ (colored). Solid lines and dashed lines represent the upper-limit ($t_{\rm lens}$) and lower-limit ($t_p$) on the timescale of plasma lensing, respectively.
}
 \label{fig: allowed lens parmeters}
\end{figure}

The difference in geometrical distance between
the wavefronts arises from ray paths outside the fireball plasma: $\overline{\rm OL_1}-\overline{\rm OL_0}=r_{\rm lens}\left[1-\cos\left(\Delta \theta/2\right)\right]$, where $\overline{\rm OL_0}$ is the path length of the closest approach from the lens to the observer. This results in the geometrical phase change due to the source motion as
\begin{eqnarray}
\label{eq: Delta phi_g}
\Delta \Phi_{g}=\frac{\nu\,(\overline{\rm OL_1}-\overline{\rm OL_0})}{c}= \nu\left(t_{\rm lens}-t_{p}\right)\xi(\Delta \theta),
\end{eqnarray}
where $\xi(x)\equiv 1-\cos\left(x/2\right)$.
Meanwhile, the dispersive phase change is expressed as
\begin{eqnarray}
\Delta\Phi_{\rm DM}(u)=-\frac{k_{\rm DM}}{\nu}\Delta{\rm DM}(u),
\end{eqnarray}
where $k_{\rm DM}=e^2/(2\pi m_e c)\sim4148.808\ {\rm s\ pc^{-1}\ cm^{-3}\ MHz^2}$is the dispersion constant and $\Delta{\rm DM}=\int n_e dz$  the excess of electron column density along the ray path.
In the case of fireball plasma lens, the electron  density evolution in the coasting phase is estimated as (equation [\ref{eq: n_e in coasting phase}) in appendix \ref{sec: appendix}]
\begin{eqnarray}
n_e(t)=2.3\times10^9  \ r_{i,5}^{-3/4}\,E_{{\rm fb},40}^{1/2}\,t^{-2}\ \ {\rm cm^{-3}},
\end{eqnarray}
where $t$ is the time from the burst onset, in units of seconds.
The increase in the dispersion measure along each ray path is
\begin{eqnarray}
\Delta{\rm DM}&=&\int_{t_{p}}^{t_{\rm lens}} n_e(t^{\prime}) \,c\,dt^{\prime}\\%
&\sim&20 \ \ r_{i,5}^{-3/4}\,E_{{\rm fb},40}^{1/2}\left(t_p^{-1}-t_{\rm lens}^{-1}\right)\ \ {\rm pc \ cm^{-3}}
\end{eqnarray}
Therefore, the dispersive phase change at $\sim$GHz frequencies is
\begin{eqnarray}
\label{eq: Delta phi_DM}
\Delta \Phi_{\rm DM}=8.3\times10^{7} \ \,\nu_9^{-1} \, r_{i,5}^{-3/4}\,E_{{\rm fb},40}^{1/2}\left(t_p^{-1}-t_{\rm lens}^{-1}\right).
\end{eqnarray}
Equating equation (\ref{eq: Delta phi_g}) and (\ref{eq: Delta phi_DM}), we obtain $t_{\rm lens}$ as a function of the pitch angle $\Delta \theta$
\begin{eqnarray}
t_{\rm lens}
\sim1.8\times10^{-4}\,\nu_9^{-1} \, r_{i,5}^{1/2}\,\left(\frac{t_p}{\tau_{\rm rec}}\right)^{-1}\,\xi(\Delta \theta)^{-1}\ \ {\rm sec}.
\end{eqnarray}
The allowed parameter space for typical $t_p$ are shown in figure \ref{fig: allowed lens parmeters}. This indicates that we can expect a strong lensing event on the order of $\gtrsim 10^3$ sec after the burst onset only when $\Delta \theta\lesssim10^{-3}$.
The pulse width for typical radio pulsars is roughly $1$--$10$ \% of the spin period, so we cannot expect to find plasma lensing for ordinary radio pulses but for a radio pulse with sub-millisecond structures.
Other possible candidates for strong plasma lensing are ultrashort duration pulses with widths of order sub-millimsecond ($\sim P\Delta\theta$), e.g., giant radio pulses in the Crab pulsar and FRBs.

In addition to the model prediction presented above, the following prediction can be made based on our hypothesis.
The energy distribution of short bursts is known to be a single power-law \citep{Cheng1996,Gogus1999,Gotz2006,Nakagawa2009}.
This implies that the number of bursts should dramatically increase with decreasing energy.
In reality, however, bursts at the faint end of the energy distribution remain undetected owing to insufficient sensitivity of current X-ray detectors. Since our model predicts the association of short bursts with radio suppression, there is a possibility that radio suppression could be used as a tracer of the short burst. Even for an extremely low-energy fireball with $E_{\rm fb}\sim 10^{34}$ erg,
the corresponding radio shut-off time scale is $\tau_{\rm rec}\sim 30\, r_{i,3}^{-7/8}\,(E_{\rm fb}/10^{34} {\rm erg})^{1/2}\,\nu_{9}^{-1}$ s, which is much longer than the typical spin period of magnetars $\sim{\cal O}(1$ s) and thus readily resolved in the radio light curve.
Therefore, the radio monitoring of known radio pulsars or magnetars for which radio pulsations are confirmed would enable us to constrain the burst rate at lower energies. Such unresolved bursts may significantly contribute to the soft thermal component of persistent X-ray emissions from magnetars \citep{TD1996,Lyutikov2003,Nakagawa2018}. The spectral similarities between short bursts and the persistent emission during the outburst phase also support this possibility \citep{Enoto2012}.

The Galactic search for pulsars with the Square Kilometer Array (SKA) is expected to yield $\sim20,000$ new pulsars \citep{Keane2015}. Statistically, a small but significant fraction (say 1 \%) of these should be  radio pulsars with magnetar activities, which could be used to test our hypothesis through simultaneous radio and X-ray observations in the near future.

\section*{Acknowledgements}
% Entry for the table of contents, for this guide only
We wish to thank Kazumi Kashiyama, Kohta Murase, Shuta J. Tanaka, Andrey Timokhin and Tomonori Totani for useful discusssion. We thank the referee Cong Yu for valuable comments which improved the quality of the manuscript.
SY is grateful to Tomonori Totani for his continuous encouragement. SY also thanks Aaron Bell, John Livingston and Akina Shimizu for providing helpful suggestions. SY was supported by Research Fellowship of Japan Society for the
Promotion of Science (JSPS) (No. {\rm 17J04010}). This work was also supported by JSPS KAKENHI Grant Number
{\rm 16J06773} (SK), {\rm 15K05069} (TT), {\rm 17H01116} (TT), {\rm 16H02198} (TE) and {\rm 18H01246} (SK, TT and TE).

\addcontentsline{toc}{section}{Acknowledgements}

%%%%%%%%%%%%%%%%%%%%%%%%%%%%%%%%%%%%%%%%%%%%%%%%%%



\bibliographystyle{mnras}
%\bibliography{draft} % if your bibtex file is called example.bib

% This method is tedious and prone to error if you have lots of refe
%\begin{thebibliography}{}
\input{draft.bbl}
%\end{thebibliography}

%%%%%%%%%%%%%%%%%%%%%%%%%%%%%%%%%%%%%%%%%%%%%%%%%%

%%%%%%%%%%%%%%%%% APPENDICES %%%%%%%%%%%%%%%%%%%%%

\appendix

\section{Relativistic plasma flows}
\label{sec: appendix}
\subsection{Dynamical pair equations}
A sudden release of pure energy into a relatively compact region around the source generate a so-called ``fireball".
We treat the fireball as a spherically expanding relativistic fluid composed of $e^{+}e^{-}$ pairs and $\gamma$ photons (``$e^{\pm}\gamma$ fireball"). Photons can be regarded as a relativistic fluid, since they are strongly coupled with pairs due to the extremely small mean free path between collisions. The conservation of energy and momentum for a steady hydrodynamical flow in spherical symmetry read
\begin{eqnarray}
	\label{eq: momentum consv.}
     \frac{1}{r^2}\frac{d}{dr}\left\{r^2(U + P)\,\Gamma^2 \beta \right\}&=& G^{0} ,\\
     \label{eq: energy consv.}
     \frac{1}{r^2}\frac{d}{dr}\left\{r^2(U+P)\,\Gamma^2 \beta^2\right\}+\frac{dP}{dr}&=& G^{1} ,
\end{eqnarray}
where $U = U_e + U_r$ is the total energy density, and $P=P_e+P_r$ being the total pressure. Subscripts ``$e$" and ``$r$" denote the plasma term and the radiation term, respectively. All the quantities are measured in the fluid rest-frame. $\beta$ and $\Gamma$ are the dimensionless three-velocity and bulk Lorentz factor defined as
$\Gamma\equiv(1-\beta^2)^{-1/2}$. $G^{\mu}$ is the radiation four-force density \citep{Mihalas1984}. Under the optically thick condition ($\tau\gg 1$), there is no radiation flux ($G^{0}=G^{1}=0$), and the radiation field stays close to the blackbody spectrum since the photons experience so many collisions that they inevitably thermalize before reaching the photosphere. Therefore $U_r = aT^4=3P_r$, where $T$ is the temperature of the fluid.
The evolution of the electron number density is tracked by the Boltzmann's equation integrated over the momentum phase space:
\begin{eqnarray}
     \label{eq: Boltzmann eq}
     \frac{1}{r^2}\frac{d}{dr}\left(r^2n_e\Gamma \beta\right)= -\langle\sigma_{\rm ann} v\rangle\left(n_e^2-n_{e,{\rm eq}}^2\right).
\end{eqnarray}
The right-hand side of equation (\ref{eq: Boltzmann eq}) represents the net pair creation rate ($\dot{n}_e$) due to collisions in $e^{+}+e^{-}\rightleftharpoons\gamma+\gamma^{\prime}$ interaction. The pair annihilation cross-section is almost constant for $kT < m_ec^2$ being approximately $\langle\sigma_{\rm ann}v\rangle\approx \pi r_e^2$ with $r_e$ the classical electron radius \citep{Svensson1982}.

\subsection{Evolution}
The evolution of $e^{\pm}\gamma$ fireball including $n_e$ evolution has been studied in great detail by \citet{Grimsrud1998}, and later developed by a number of authors (e.g., \citealt{Iwamoto2002,Nakar2005,Li2008}). Here we follow their formulation.
We assume that the fireball starts with initial conditions, i.e., a Lorentz factor $\Gamma_i$, temperature $T_i$ and radius $r_i$. Combining the equation of state for a relativistic gas $U=3P$, equation (\ref{eq: momentum consv.}) and (\ref{eq: energy consv.}) reduce to a set of useful scaling laws that describe the radial evolution of Lorentz factor and temperature (\citealt{Paczynski1986,Goodman1986}):
\begin{eqnarray}
\label{eq: FB evol}
\Gamma\approx\Gamma_i(r/r_i) , \ \ T\approx T_i(r_i/r).
\end{eqnarray}
As will be shown below, the evolution of the fireball falls into five phases, characterized by the corresponding radii that determine the physical state of the fireball.

\subsubsection{Phase I ($\tau\gg1$ and $T\gtrsim m_e$)}
Initially, the fireball temperature is expected to large ($T\gg m_e$) so that the pair plasma is in equilibrium state $n_e=n_{e,\rm eq}$, where $n_{e,{\rm eq}}$ is described in equation (\ref{eq: n_e,eq}). The fireball is initially confined at rest and immediately start to expand at a speed close to $c$, while the temperature decreases. Let $r=r_m$ the radius at which the temperature equals to the electron rest mass energy. Then we obtain
\begin{eqnarray}
\frac{r_m}{r_i}\approx\frac{T_i}{m_e},
\end{eqnarray}
simply because $Tr=$ const. from equation (\ref{eq: FB evol}).
The density evolution for $r>r_m$ is
\begin{eqnarray}
\label{eq: phase I}
n_{e}(r)=\frac{2}{(2\pi)^{3/2}}\,\lambda_C^{-3}\,\left(\frac{r_m}{r}\right)^{3/2}\,e^{-r/r_m}
\end{eqnarray}
as long as the pair equilibrium is met.

\subsubsection{Phase II ($\tau\gg1$ and $T< m_e$)}
After the phase I, $n_e$ soon starts to deviate from $n_{e,\rm eq}$, since $n_{e,\rm eq}$ decays exponentially with decreasing $T$. Let us assume a small deviation $\delta n_{e,\rm eq}$ from $n_e=n_{e,\rm eq}$ so that $n_e^2 -n_{e,{\rm eq}}^2\approx 2\,n_{e,{\rm eq}}\,\delta n_e$. Then equation (\ref{eq: Boltzmann eq}) yields
\begin{eqnarray}
    \delta n_e &\approx&-\frac{1}{2r^2\langle\sigma_{\rm ann}v\rangle n_{e,{\rm eq}}}\frac{d}{dr}\left(r^2n_{e,{\rm eq}} \Gamma\beta\right)
    \nonumber \\%
    &\approx&-\frac{\Gamma}{2r\langle\sigma_{\rm ann}v\rangle}\left[3+\frac{d\ln{n_{e,{\rm eq}}}}{d\ln{r}}\right]\nonumber \\%
    &\approx&\frac{\Gamma}{2r\langle\sigma_{\rm ann}v\rangle}\frac{m_e}{T},
\end{eqnarray}
where we make use of $\Gamma\propto r$ in the second deformation and  $d\ln{n_{e,\rm eq}}/d\ln{T}\sim m_e/T$ in the third. Let $r_{\pm}$ ($T_{\pm}$) the radius (temperature) at which the deviation grows to the same order as the equilibrium density ($n_{e,\rm eq}=\delta n_e$). Then we obtain the following equation
\begin{eqnarray}
 \frac{2}{(2\pi)^{3/2}}\lambda_C^{-3}\left(T_{\pm}/m_e\right)^{3/2}\,e^{-m_e/T_{\pm}}=\frac{\Gamma_{\pm}}{2\pi r_e^2 \,r_{\pm}}\frac{m_e}{T_{\pm}}.
\end{eqnarray}
Using $T_{\pm}r_{\pm}\approx m_er_m$ and $\Gamma_{\pm}/r_{\pm}\approx \Gamma_i/r_i$, this reduces to
\begin{eqnarray}
\label{eq: r_pm}
\frac{r_{\pm}}{r_m}=\ln{\left\{\sqrt{2/\pi}\lambda_C^{-3}r_e^2\,r_i/\Gamma_i\right\}}-{\frac{5}{2}}\ln{\left(\frac{r_{\pm}}{r_m}\right)}
\end{eqnarray}
It can be seen that $r_{\pm}/r_m$ weakly depends on $r_i$. Evaluating equation (\ref{eq: r_pm}) numerically, we obtain an analytic fitting formula:
\begin{eqnarray}
\label{eq: analytic r_pm/r_m}
\frac{r_{\pm}}{r_m}\approx13.4+\log{\left(r_{i,5}\right)}.
\end{eqnarray}
Thus we get $r_{\pm}/r_m\sim13$ for a reference initial radius $r_i=10^5$ cm \footnote{\citet{Grimsrud1998} originally obtained $r_{\pm}/r_m\sim33$ for $r_i=10^6$ cm, which seems to an overestimate because they simply neglect $\ln{\left(r_{\pm}/r_m\right)}$ appearing in the  right-hand side of equation (\ref{eq: r_pm}).}, and the corresponding temperature $T_{\pm}=m_e(r_m/r_{\pm})\sim0.08\,m_e$ (weakly dependent on $r_i$). Consequently, the electron number density at $r=r_{\pm}$ is estimated as
\begin{eqnarray}
\label{eq: phase II}
n_{e}(r_{\pm})&=&n_{e,\rm eq}(r_{\pm})+\delta n_{e}(r_{\pm})\approx 2\,\delta n_{e}(r_{\pm})
\nonumber \\%
&\sim&\frac{1}{\pi r_e^2}\left(\frac{\Gamma_i}{r_i}\right)\left(\frac{r_{\pm}}{r_m}\right)
\nonumber \\%
&\sim&5.2\times10^{20}\,\Gamma_i \,r_{i,5}^{-1} \ \ {\rm cm^{-3}},
\end{eqnarray}
where we have used equation (\ref{eq: analytic r_pm/r_m}) in the last derivation.

\subsubsection{Phase III ($\tau\sim1$ and $T\lesssim m_e$)}
For $r>r_{\pm}$, pairs have already deviated greatly from the equilibrium state ($n_e\gg n_{e,\rm eq}$) and the pair equation (\ref{eq: Boltzmann eq}) becomes simply
\begin{eqnarray}
     \frac{1}{r^2}\frac{d}{dr}\left(r^3n_e \right)= -\pi r_e^2 \frac{r_{\pm}}{\Gamma_{\pm}}n_e^2,
\end{eqnarray}
where we have used $\Gamma=\Gamma_{\pm}(r/r_{\pm})$. Namely, the pair annihilation dominates the pair creation since the number of high-energy photons decrease. This can be solved analytically:
\begin{eqnarray}
\label{eq: phase III to IV}
    n_e(r)=\frac{n_e(r_{\pm})}{1+(1/3)(r_{\pm}/r_m)[1-(r_{\pm}/r)^3]}\left(\frac{r_{\pm}}{r}\right)^3.
\end{eqnarray}
Then we can calculate the optical depth to the electron scattering as
\begin{eqnarray}
\tau(r)&=&\int_{r}^{\infty}2n_e(s)\sigma_T(1-\beta)\Gamma \,ds
\nonumber \\%
&=&-\frac{8}{3}\ln{\left\{1-\frac{(1/3)(r_{\pm}/r_m)(r_{\pm}/r)^3}{1+(1/3)(r_{\pm}/r_m)}\right\}}
\end{eqnarray}
where $\sigma_T=8\pi/3\,r_e^2$ is the Thomson cross-section.
We define the photospheric radius $r_{ph}$ at which the optical depth becomes unity [$\tau(r_{ph})=1$]. Then we get
\begin{eqnarray}
\frac{r_{ph}}{r_{\pm}}=\left\{\left(1-e^{-3/8}\right)\left(3\,\frac{r_m}{r_{\pm}}+1\right)\right\}^{-1/3}\sim1.4,
\end{eqnarray}
with the corresponding temperature $T_{ph}\sim T_{\pm}(r_{\pm}/r_{ph})\sim0.06\,m_e$.

\subsubsection{Phase IV ($\tau\lesssim1$ and $T\ll m_e$)}
When $\tau\lesssim1$, photons start to stream freely, and the single fluid approximation [equation (\ref{eq: momentum consv.}) and (\ref{eq: energy consv.})] does not hold anymore. Nevertheless, pairs continue to accelerate up to the certain point where the escaping radiation could no longer supply enough energy to pairs.
The Lorentz factor of pairs ($\Gamma$) evolves as
\begin{eqnarray}
     \frac{d\Gamma}{dr}=\frac{\sigma_TF_r}{m_e c^3},
\end{eqnarray}
where $F_r$ is the photon energy flux in the local rest frame of pairs. We can relate the photon energy flux in the rest frame of pairs to the photon internal energy in the rest frame of photons by a simple Lorentz transformation:
\begin{eqnarray}
F_r=\Gamma_{\rm rel}^2\beta_{\rm rel}c\,(U_r^{\prime}+P_r^{\prime}),
\end{eqnarray}
where $U_r^{\prime}$ and $P_r^{\prime}(=U_r^{\prime}/3)$ are the radiation energy density and pressure in the rest frame of photons, respectively.
Since the photon energy density in the rest frame of free streaming photons can be approximated by a blackbody, we can write as $U_r^{\prime}=aT_r^4$, where $T_r$ is the photon temperature in its rest frame \citep{Li2008}. $\Gamma_{\rm rel} $ is the Lorentz factor for the relative velocity $\beta_{\rm rel}$ of photons ($\beta_r$) with respect to the pairs ($\beta$), which can be written as
\begin{eqnarray}
\Gamma_{\rm rel}=\Gamma\Gamma_{r}(1-\beta\beta_{r})\sim\frac{1}{2}\left(\frac{\Gamma_{r}}{\Gamma}+\frac{\Gamma}{\Gamma_{r}}\right),
\end{eqnarray}
where $\Gamma_r$ is the Lorentz factor of photons. Note that photons do not completely decouple from the pairs at this stage and thus $\Gamma_r$ has a finite value.
Then we arrive at
\begin{eqnarray}
\frac{d\Gamma}{dr}=\frac{aT_r^4}{3}\left[\left(\frac{\Gamma_{r}}{\Gamma}\right)^2-\left(\frac{\Gamma}{\Gamma_{r}}\right)^2\right]\frac{\sigma_T}{m_e c^2}
\end{eqnarray}
Combining $T_r=T_i (r_i/r)$ and $\Gamma_r=\Gamma_i(r/r_i)$, the asymptotic Lorentz factor of pairs is obtained as \citep{Li2008}
\begin{eqnarray}
\label{eq: coasting Lorentz factor}
\Gamma_{\infty}\equiv \lim_{r \to \infty} \Gamma \sim1.46\left(\frac{3m_ec^2}{r_{i}aT_{i}^4\sigma_T}\right)^{-1/4}\nonumber \\%
\sim1.1\times10^3 \,\Gamma_i\,r_{i,5}^{1/4}\,\left(\frac{T_i}{m_e}\right)
\end{eqnarray}
Conversely, the radius $r=r_{\infty}$ at which $\Gamma=\Gamma_{\infty}$ is estimated as
$r_{\infty}\approx(\Gamma_{\infty}/\Gamma_i)r_i$. Relating $r_{\infty}$ with $r_{ph}\sim18\,(T_i/m_e)\,r_i$, we finally obtain
\begin{eqnarray}
\frac{r_{\infty}}{r_{ph}}\sim61\,r_{i,5}^{1/4}\,\left(\frac{T_i}{m_e}\right)
\end{eqnarray}
The electron number density at $r=r_{\infty}$ is
$n_e(r_{\infty})\sim3.3\times10^{-7}\,r_{i,5}^{-3/4}\,n_e(r_{\pm})\sim1.7\times10^{14}\,r_{i,5}^{-7/4} \ \ {\rm cm^{-3}}$.

\subsubsection{Phase V ($\tau\ll1$ and $T\ll m_e$)}
For $r>r_{\infty}$, the pair annihilation does not occur any more due to the small pair number density, and thus the total number of pairs preserves. Neglecting the right-hand side of equation (\ref{eq: Boltzmann eq}), we obtain $n_er^2=$ const.
Namely, the evolution of the electron number density is
\begin{eqnarray}
\label{eq: phase V}
\label{eq: n_e in coasting phase}
    n_e(r)=n_e(r_{\infty})\left(\frac{r_{\infty}}{r}\right)^2.
\end{eqnarray}

\subsection{Baryon loaded fireball}
\label{subsec: Baryon loaded fireball}
If the initial fireball forms in the vicinity of the neutron star surface, it is expected that some amount of baryons should be contaminated, which could break the equality of the pair number density ($n_{e^{-}}>n_{e^{+}}$) and might affect the late time evolution. Conservation of baryon number and energy reads
\begin{eqnarray}
\label{eq: M_dot and E_dot}
    \dot{M}&=&4\pi r^2Am_p\,n\,\Gamma \beta = const,\\
    L&=&4\pi r^2(U+P)\Gamma^2\beta=const,
\end{eqnarray}
where $n$ is the baryon number density with mass number $A$ (and  atomic number $Z$) and $m_p$ being the proton mass. We assume the charge neutrality $n_{e^{-}}=n_{e^{+}}+Zn$ and introduce a dimensionless entropy
\begin{eqnarray}
\label{eq: eta}
    \eta\equiv\frac{L}{\dot{M}},
\end{eqnarray}
which represents the radiation-to-baryon ratio.
We can see that the adiabatic evolution ($\Gamma\propto r$ and $T\propto 1/r$) breaks up when the kinetic energy begins to dominate the radiation energy. This transition takes place when $U+P\sim Am_pn$ with a corresponding radius $r_M=\eta \,(r_i/\Gamma_i)$, above which the Lorentz factor stays constant ($\Gamma_{\infty}=\eta$).
A critical value of $\eta$ is obtained as
\begin{eqnarray}
    \eta_{c}\sim200\left(\frac{Z}{A}\right)^{1/4}r_{i,5}^{1/4}\,\Gamma_{i}^{3/4}\,\frac{T_i}{m_e},
\end{eqnarray}
by simply setting $r_M=r_{ph}$\footnote{Note that, the optical depth is approximated as $\tau\approx Zn\sigma_Tr/\Gamma$, taking into account baryon-associated electrons.}.

In the case of extremely heavy baryon loading $\eta\lesssim\eta_{c}$, the number density of positrons becomes negligible compared to that of both electrons and baryons (hence $n_e\sim Zn$).
Setting $(U+P)_i\sim aT_i^4$ in a set of equations (\ref{eq: M_dot and E_dot})--(\ref{eq: eta}), the radial evolution of the electron number density may be estimated as
\begin{eqnarray}
    n_e\sim \frac{aT_i^4\Gamma_i}{m_p} \left(\frac{Z}{A}\right)\times\begin{cases}
    \eta^{-1}\left(r/r_i\right)^{-3} & (r<r_M) \\
    \eta^{-2}\left(r/r_i\right)^{-2} & (r>r_M),
  \end{cases}
\end{eqnarray}
Remarkably, the observed plasma frequency $\nu_p\propto \Gamma n_e^{1/2}$ does not depend on $\eta$ at $r>r_M$ since $n_e\propto\eta^{-2}$ and $\Gamma=\eta$.

Meanwhile, for a moderate baryon loading $\eta\gg\eta_{c}$, the coasting Lorentz factor can be estimated in much the same manner as we employed with baryon-free fireball.
We define the effective electron mass ${\tilde{m}}_e$ as
\begin{eqnarray}
    {\tilde{m}}_e=\frac{2m_en_e+Am_pn}{2n_e}\sim m_e+\frac{Am_p}{2Z}.
\end{eqnarray}
Here we assume the number density of electrons and positrons are nearly equal outside the photospheric radius. By replacing $m_e$ with ${\tilde{m}}_e$ in equation (\ref{eq: coasting Lorentz factor}), the coasting Lorentz factor $\Gamma_{\infty}$ is found to reduce at most by a factor of $(Am_p/2Zm_e)^{1/4}\sim6\,(A/Z)^{1/4}$ compared to the baryon-free case. The inequalty between electron/positron number density does not significantly change $r_{\rm eq}$ and $r_{ph}$ throughout $\eta>\eta_c$ \citep{Grimsrud1998}.

Therefore, we conclude that the evolution of electron number density conserves as long as the baryon contamination is small ($\eta\gg\eta_c$). Given the heavy baryon loading, the density at the coasting phase (phase V) would reduce by a factor of $6\,(A/Z)^{1/4}$ at most, since $n_e(r)\propto r_{\infty}^{-1}\propto \Gamma_{\infty}^{-1}$.

%%%%%%%%%%%%%%%%%%%%%%%%%%%%%%%%%%%%%%%%%%%%%%%%%%

% Don't change these lines
\bsp	% typesetting comment
\label{lastpage}
\end{document}